\pdfoutput=1
\documentclass{aa}
\usepackage[varg]{txfonts}
\usepackage{natbib}
\usepackage{longtable}
\usepackage{graphicx}

\begin{document}

\title{Clustering of galaxies around GRB sight-lines}


\author{Vladimir Sudilovsky \inst{\ref{MPE}}
\and Jochen Greiner \inst{\ref{MPE}}
\and Arne Rau \inst{\ref{MPE}}
\and Mara Salvato \inst{\ref{MPE}}
\and Sandra Savaglio \inst{\ref{MPE}}
\and Susanna D. Vergani \inst{\ref{GEPI},\ref{INAF}}
\and P. Schady \inst{\ref{MPE}}
\and Jonny Elliott \inst{\ref{MPE}}
\and T. Kr{\"u}hler \inst{\ref{DARK}}
\and D. A. Kann \inst{\ref{TLS}}
\and Sylvio Klose \inst{\ref{TLS}}
\and Andrea Rossi \inst{\ref{TLS}}
\and Robert Filgas \inst{\ref{IEAP}}
\and Sebastian Schmidl \inst{\ref{TLS}}
}

\institute{Max-Planck-Institut f\"ur extraterrestrische Physik, Giessenbachstrasse, D-85748 Garching bei M\"unchen,
Germany \label{MPE}
\and GEPI, Observatoire de Paris, CNRS, Univ. Paris Diderot, 5 Place Jules Jannsen, F-92195, Meudon, France \label{GEPI}
\and INAF, Osservatorio Astronomico di Brera, via E. Bianchi 46, 23807 Merate, Italy \label{INAF}
\and Dark Cosmology Centre, Niels Bohr Institute, University of Copenhagen, Juliane Maries Vej 30, 2100 Copenhagen, Denmark
\label{DARK}
\and Th\"uringer Landessternwarte Tautenburg, Sternwarte 5, 07778 Tautenburg, Germany \label{TLS}
\and Institute of Experimental and Applied Physics, Czech Technical University in Prague, Horsk\'a 3a/22, 128 00 Prague 2, Czech
Republic \label{IEAP}
}
\date{Recieved x / Accepted x}

\abstract{There is evidence of an overdensity of strong intervening MgII absorption line systems
distributed along the lines of sight towards GRB afterglows relative to quasar sight-lines. If this excess is real, one should
also
expect an overdensity of field galaxies around GRB
sight-lines, as strong MgII tends to trace these sources. In this work, we test this expectation by calculating the two
point
angular correlation function of galaxies within 120$^{\prime\prime}$ ($\sim470~h_{71}^{-1}~\mathrm{Kpc}$ at
$\langle z\rangle \sim0.4$) of GRB afterglows. We compare the Gamma-ray burst Optical and
Near-infrared Detector (GROND) GRB afterglow sample -- one of the
largest and most homogeneous samples of GRB fields -- with galaxies and AGN found in the
COSMOS-30 photometric catalog. We find no significant signal of anomalous clustering of galaxies at an estimated median redshift
of $z\sim0.3$ around GRB sight-lines,
down to $K_{\mathrm{AB}}<19.3$. This result is contrary to
the expectations from the MgII excess derived from GRB afterglow spectroscopy, although many confirmed galaxy counterparts to MgII
absorbers may be too faint to detect in our sample -- especially those at $z>1$. We note that the addition of higher sensitivity
Spitzer IRAC or HST WFC3 data for even a subset of our
sample would increase this survey's depth by several orders of magnitude, simultaneously increasing statistics and enabling the
investigation of a much larger redshift space.}


\authorrunning{Sudilovsky et al.}
\maketitle

\section{Introduction} \label{section:introduction} 
Gamma-ray bursts (GRBs) have proven to be powerful tools for studying the high redshift universe. Their afterglows pinpoint star
forming galaxies that would otherwise be exceedingly difficult to discover. Using bright transients such as GRBs to study the
Universe introduces completely different selection criteria from standard surveying techniques. Long duration GRB (in this
work, the term ``GRBs'' implies long-duration GRBs, unless explicitly stated otherwise) afterglows have been used to study the
star formation and metallicity evolution of the Universe \citep[see
e.g.][]{SAVAGLIO2009,JAKOBSSON2005,PROCHASKA2006,SALVATERRA2012}. The host galaxies of $z>5$ GRB afterglows have been
studied by \citet{TANVIR2012,BASA2012}. The upper limits on the non-detections that the authors derive suggest that the galaxy
luminosity function evolves rapidly at these higher redshifts. In general, GRBs seem to be clear tracers of star formation
\citep[see e.g.][]{LEFLOCH2006,BUTLER2010}, however it is unclear the extent to which selection effects against more dusty
and metal-rich galaxies may affect surveys\citep{KRUEHLER2011,ELLIOTT2012,PERLEY2013}.

In addition to studies of the GRB host galaxies, GRB afterglows themselves offer a brief and bright glimpse into the 
high redshift universe. Intervening absorption line systems can be detected in the same manner employed during
the past decades with quasar spectroscopy. One such absorption line system is MgII, which is easily detected in moderate S/N
spectra at $\lambda_{\mathrm{rest}} \sim 2800$~\AA. Because of its strong absorption
and ease of identification owing to the fact that it is an absorption doublet, MgII has been used extensively as a tracer of
galaxies, galactic outflows, and chemical evolution. MgII itself is
coincident with a wide range of neutral hydrogen column densities from $N_{HI} \sim 10^{16}-10^{22} \mathrm{cm}^{-2}$
\citep{CHURCHILL2000}, though there is strong evidence that MgII equivalent width (EW) is directly correlated with higher hydrogen
columns and thus smaller impact parameters to the bulk star forming region of
galaxies \citep{STEIDEL1995,BOUCHE2006,KACPRZAK2011B,BORDOLOI2011}. Indeed, \citet{BORDOLOI2012} showed that MgII absorbers
follow a bi-modal spatial distribution, wherein at impact parameters smaller than $d \sim 40$~Kpc, MgII is associated with cool
star forming regions, and at $d > 40$~Kpc MgII tends to be uniformly distributed around galaxies, perhaps suggesting large scale
outflows. There is no evidence to suggest that this behavior evolves with redshift, at least in the interval $0.4 \le z
\le 2$. It is, however worth noting that \citet{MATEJEK2012} have studied the
$2<z<6$ regime using infrared spectra of quasars. The authors find that the association of MgII with damped Lyman-$\alpha$
systems (DLAs) strengthens towards higher redshifts, even though the overall taxonomy of these absorbers as defined in
\citet{CHURCHILL2000} does not evolve. There is evidence that the redshift evolution in the number density of MgII absorbers
follows the cosmic star formation rate \citep{ZHU2012,MATEJEK2012A}, though the effects of observational bias may still play an
important role \citep{LOPEZ2012}.

\citet{PROCHTER2006} compared the incidence of strong MgII absorbers in quasar
sight-lines with those found in GRB afterglows. In this context, ``strong'' is defined as absorption systems whose rest
frame equivalent width of the $\lambda 2796$ feature is $W_0 \ge 1.0$ \AA. The authors found a factor of $\sim4$ excess in
the number density of these strong 
intervening $0.4<z<2.0$ MgII systems in GRB sight-lines relative to those found in quasars sight-lines. More
recently, \citet{VERGANI2009}
have
confirmed a factor of $\sim2$
excess with a much larger sample. Besides increasing the statistical significance of the MgII discrepancy, they
found that the properties of weak $0.3 \le W_0 \le 1.0$~\AA~intervening MgII systems are statistically identical
between the two
types of sight-lines. Interestingly, the abundance of CIV -- a
higher ionization absorption line system -- also does not show
an overdensity \citep{SUDILOVSKY2007,TEJOS2007}. Many authors have examined possible solutions to the observed discrepancy, and
agree that a satisfactory explanation does not yet exist
\citep{FRANK2007,SUDILOVSKY2009,CUCCHIARA2009,VERGANI2009,KANN2010,WYITHE2011,RAPOPORT2012}. A
comprehensive overview of these possible solutions is given in
\citet{PORCIANI2007}. It is especially intriguing that the amplitude of the discrepancy seemingly depends on the
resolution of the afterglow spectra suggesting either that rapid response high resolution
spectroscopy may introduce a not yet understood observational bias in any sample study of GRB afterglows, or that the MgII
discrepancy is a statistical fluke \citep{CUCCHIARA2012_ARXIV}.

Objects near the line of sight to a GRB afterglow are unaffected by light from the event
after the afterglow has faded, allowing detailed follow-up of the field to extremely high angular resolutions. The study of
sources near quasars are reliant on either extremely high resolution imaging on intrinsically dim quasars, Lyman-$\alpha$  
imaging, or very low redshift quasars. Galaxies that give rise to absorption line systems in GRB afterglow spectra have been
directly imaged and studied, although the number of robust associations is still extremely small.
\citet{CHEN2009} find that additional galaxies are found at very close angular distances to GRB host galaxies whose afterglows
exhibited strong MgII absorption, though it is unclear what fraction were associated with the absorption line systems.
\citet{SCHULZE2012} proposed galaxy counterpart candidates to absorbers, and additionally found field
galaxies with the same redshift as absorption line systems at distances of 130-161 kpc away from the sight-line. The frequency of
field galaxies that are associated with MgII absorption line systems is still unclear. However, \citet{LOPEZ2008} have shown
that strong MgII absorbers are strongly associated with galaxy clusters, and that the number density of these absorbers is much
higher relative to field galaxies. In this work, we test for any anomalous signature of clustering or overabundance
of field galaxies in GRB and quasar lines of sight.

If there is indeed a higher probability of detecting strong MgII absorption line systems in GRB afterglow than in quasar
sight-lines, one should also expect to detect more galaxies at close angular separations to GRB sight-lines. The two point
correlation function is a powerful tool to determine if objects are clustered, and, if so, what their correlation lengths are. The
angular two point correlation function is formally defined as

\begin{equation}
\begin{centering}
dP = n[1+w(\theta)]d\Omega, \label{eqn:formal_tpc}
\end{centering}
\end{equation}

where $dP$ is the probability of finding an object within a solid angle $d\Omega$ at an angular distance $\theta$
\citep{PEEBLES1980}. In general, a positive $w$ implies some enhancement of object-object grouping above a uniform random
distribution, while a negative $w$ implies some avoidance. The two point correlation function has been used to estimate
the clustering properties of galaxies, quasars, and GRBs \citep[see e.g.][]{GROTH1977,ROSS2009,BRAINERD1995}.

In this work, we measure the angular two-point correlation function between GRB afterglows and field galaxies,
and compare this quantity with galaxy-galaxy and AGN-galaxy correlations. In \S \ref{section:sample}, we present the sample that
we analyze in \S \ref{section:procedure}. We present the results of the analysis in \S \ref{section:results}, and discuss these
results in \S \ref{section:discussion}. 

\section{The sample} \label{section:sample}
The Gamma-ray burst Optical and Near-infrared Detector (GROND) is a simultaneous 7-channel imager mounted on the MPG/ESO 2.2m
telescope at La Silla, Chile \citep{GROND_TECHPAPER}. The
four optical channels are nearly identical to the SDSS $g^\prime r^\prime i^\prime z^\prime$ filters, while the infrared channels
are effectively equivalent to the 2MASS
$JHK$ filters. While the detector system was built to quickly measure photometric
redshifts of GRBs via the Lyman-$\alpha$ dropout technique, the multi-wavelength photometry provided by GROND
can potentially reveal a wealth of physical characteristics of any source \citep{GROND_TECHPAPER}.

\begin{figure}
	\resizebox{\hsize}{!}{
	\includegraphics[width=7cm]{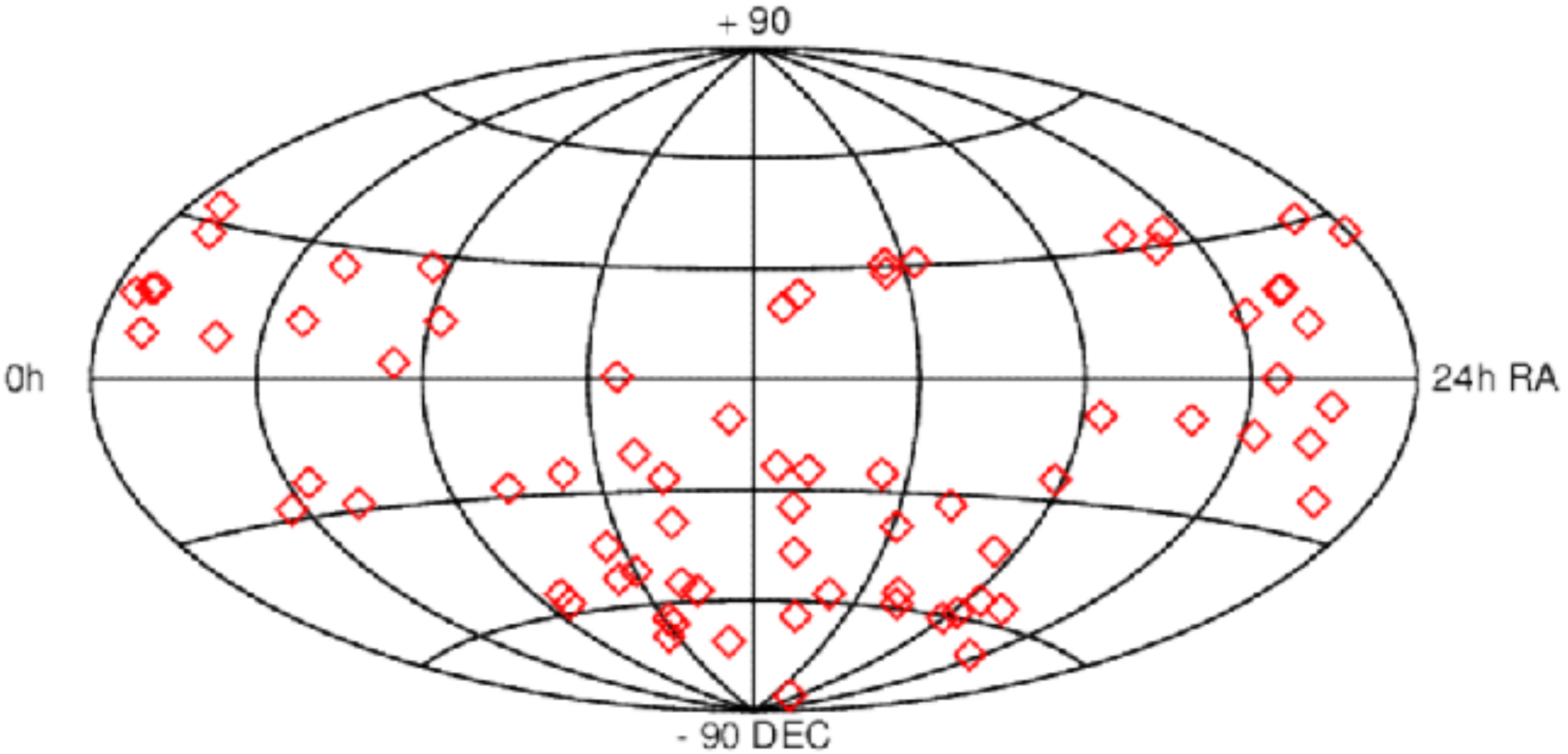}}
        \caption{\emph{Positions in equatorial coordinates of the 73 GROND GRB afterglows in the sample. Bursts within
10$^{\circ}$ of the Galactic plane
are excluded.}}
        \label{fig:map}
        \end{figure}

Our sample of GRB fields is constructed exclusively of GROND observations taken between 2007-2012. This fact ensures a high degree
of homogeneity, as no correction for cross-instrumental calibrations are required. Furthermore, the simultaneous nature of GROND
observations ensures that data on a per-field basis have not varied due to intrinsic variability or weather conditions.
Since only ephemeris, hardware failure and weather losses 
prevent GROND follow-up observations, the GROND sample of GRB afterglows has an exceptional success rate of $\sim90$\% detection
of long-duration GRBs when follow-up is possible within the first few hours after the trigger \citep{GREINER2011,KRUEHLER2011}.
We include in our sample all fields which 1) have a XRT localization of the burst, 2) Galactic
latitude
$|b| > 10^{\circ}$, 3) at least $15^{\circ}$ from the Galactic center, and 4) have been observed to a 3 $\sigma$ background
limiting magnitude of K$_{\mathrm{AB}}\sim20.4$. This limiting magnitude corresponds roughly to an integration time of 1500 and
1200
seconds in optical and NIR, respectively, among four telescope dither positions. For
reference, typical 3$\sigma$ AB
limiting magnitudes in the optical are at least 3 mag deeper than K, while J and H are 1.0 and 0.5 mag deeper, respectively. Our
final sample
consists of 73 GRB fields (see Table \ref{table:sample} and Fig. \ref{fig:map}). Since most of the bursts have no measured
redshift, we assume that our bursts follows the same redshift distribution as in the TOUGH sample \citep{JAKOBSSON2012}.

\section{Procedure} \label{section:procedure}
\subsection{Creating the source catalog} \label{subsection:creating_catalog}
To accurately measure the optimal extraction region and measure the apparent extent of sources from GROND multi-band images, we
co-add all 7 images to create a single detection
image for every field. Since the exposure time differs between the NIR and optical channels, we scale and weigh the images to
a common effective exposure. SExtractor \citep{SEXTRACTOR} is run on the detection image using relatively low source extraction
thresholds such that ~5-10\% of the sources are likely spurious detections, but still no bright sources are incorrectly de-blended
or otherwise split into multiple sources. The detection image source catalog is correlated with each individual band's source
catalog, and only those sources whose 1)
positions in each catalog are consistent to within 0\farcs5 of each other, and 2) are detected in at least 6 out of the 7 bands
(one of which is required to be a K-band detection), and 3) have an error $\sigma_{K}\le0.10$ mag are used in the final source
catalog for
that field. The latter criteria limits the overall sample to roughly $\sigma_K < 19.3$, with $\sim10\%$ of sources dimmer
than
this. The incompleteness down to $K~\sim20.0$ does not introduce any significant bias in the final results. This cross-correlation
ensures that
spurious detections are removed, and that each source has multi-wavelength
detections. The requirement that objects be strongly detected in $K$ ensures that they can be reliably seperated between galaxies
and stars.

Creating a source catalog based on a detection image in the aforementioned manner minimizes the probability that flux from
extended objects is missed due to varying spectral properties as a function of position. That is, by determining a suitable
aperture from the co-added image, one is guaranteed that the entirety of flux in an extended source is included therein if that
same aperture is then used in each individual image, as long as that source is strongly detected in each band. Additionally, if 
the source is reasonably isolated in the image, this method avoids the complications introduced by PSF matching and galaxy fitting
while still providing adequate photometry. We therefore perform aperture photometry with SExtractor in dual-mode using the
detection image as the template for each band. 

The SExtractor magnitudes are calibrated based on time-tabulated instrumental zeropoints in the case of the optical channels and
2MASS field stars in the case of the NIR. GROND zeropoints are calibrated on average once every three months based on SDSS
standard fields, and immediately after technical work is performed on the optical system. The RMS of the optical zeropoints as a
function
of time is on the order of $0.05$ mag. We expect a similar spread due to varying atmospheric conditions, as the range of seeing in
our sample is between $1.0$ and $1.6$ in $r^\prime$. In $18$ fields ($24$\%) co-incident with the SDSS, magnitudes were
calibrated directly against SDSS field stars.

Next, we fit each object's spectral energy distribution (SED) with \textit{LePHARE} \citep{LEPHARE1,LEPHARE2}. \textit{LePHARE} is
a
spectral template fitting tool based on $\chi^2$ minimization. We determine the best fit galaxy and stellar templates for each
source, using the COSMOS galaxy
templates, with emission lines and prescribed reddening and parameters therein \citep{ILBERT2009}. More specifically, we
adopted the Prevot \citep{PREVOT1984} extinction law for late type templates and the Calzetti law \citep{CALZETTI1994} and two
modifications thereof \citep{ILBERT2009} for the SB templates. The values of extinction range from $0.0 < E(B-V) < 0.5$ in
steps of $0.1$. Early type templates are not corrected for extinction as there are not well tested empirical models to describe
the dust distribution for these galaxies. In the case of
galaxy templates,
redshift is constrained to $z<2$ with a step size $\Delta z = 0.01$. The former constraint removes some degeneracy between high
and low redshift template fits, and is justified since the population of galaxies with an apparent K$_{\mathrm{AB}}<20.0$ is
negligibly
small at $z\gtrsim1.5$. We assume that
the contamination from AGN-dominated sources in our galaxy catalog is negligible ($\lesssim5$\%, as discussed later in the paper).
We separate galaxies and stars based on their best fit template
and
shape parameters. Any sources with a full-width half maximum (FWHM) or ellipticity as measured with SExtractor in $5\sigma$ excess
of the stellar values for
that field are automatically
classified as galaxies. These stellar values are determined by computing the average and standard deviation of the lowest 20\%
FWHM and ellipticity distributions, under the assumption that those ``bottom 20\%" of sources are stars. In practice, 
this usually means the FWHM and ellipticity of 5-10 stars are averaged for a given field. Besides this shape criterion, we
categorize sources with $1.5 \times \chi^2_{\mathrm{star}} > \chi^2_{\mathrm{galaxy}}$ as galaxies, where $\chi^2$ corresponds to
the best fit template for that class of source. This latter criterion has been
applied to the COSMOS field with great success \citep{SALVATO2009}. The extinction law and extinction values are to
some extent free parameters in the fitting. Therefore, a certain amount of degeneracy in redshift-color-extinction space is
expected. However, we do not expect this degeneracy to affect star/galaxy classification. This is due to the consideration of
morphological information (i.e. extended vs point-like) and the implied high redshift limit
imposed by the $K$ band magnitude
limit. 

\begin{figure}
	\resizebox{\hsize}{!}{
	\includegraphics[width=7cm]{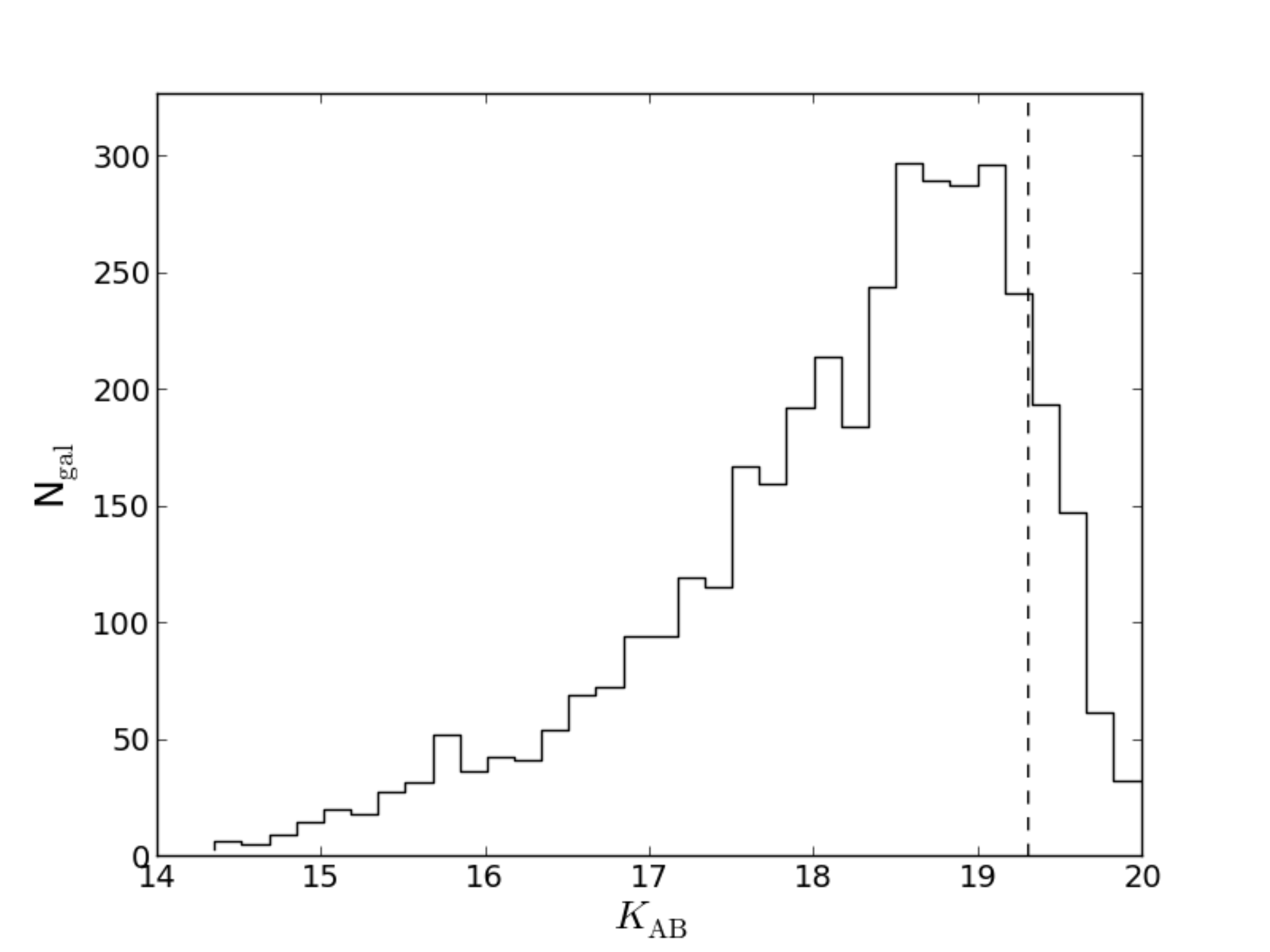}}
        \caption{\emph{K$_{\mathrm{AB}}$ magnitude distribution of the 3068 field galaxies identified in the sample. The steep
drop-off after $K~\sim19.3$ is due to the criteria that $\sigma_{K}\le0.10$ for each galaxy, and that is it detected in five other
bands. This
represents a decline in completeness of the sample. The typical $3\sigma$ limiting magnitude for an individual image as determined
from the sky background is $K_{\mathrm{AB}}\sim20.4$.}}
        \label{fig:K_distro}
        \end{figure}

We detect $3068$ galaxies and $1368$ stars that match our
criteria. The $K$-band magnitude distribution of galaxies is presented in Fig. \ref{fig:K_distro}.

\subsection{Verifying the source catalog}
        \begin{figure*}
\resizebox{\hsize}{!}{
\includegraphics[width=4.0cm]{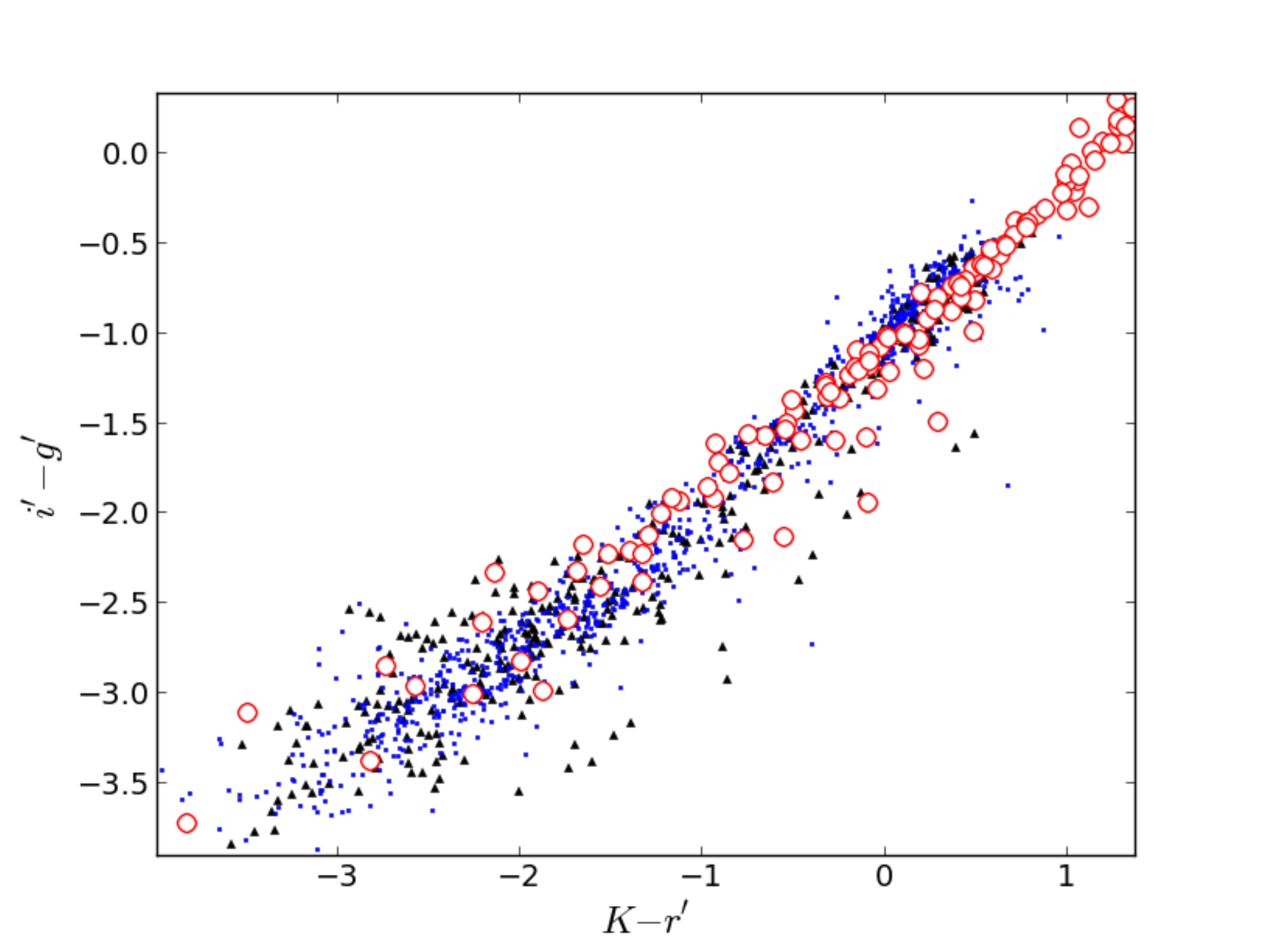}\includegraphics[width=4.0cm]{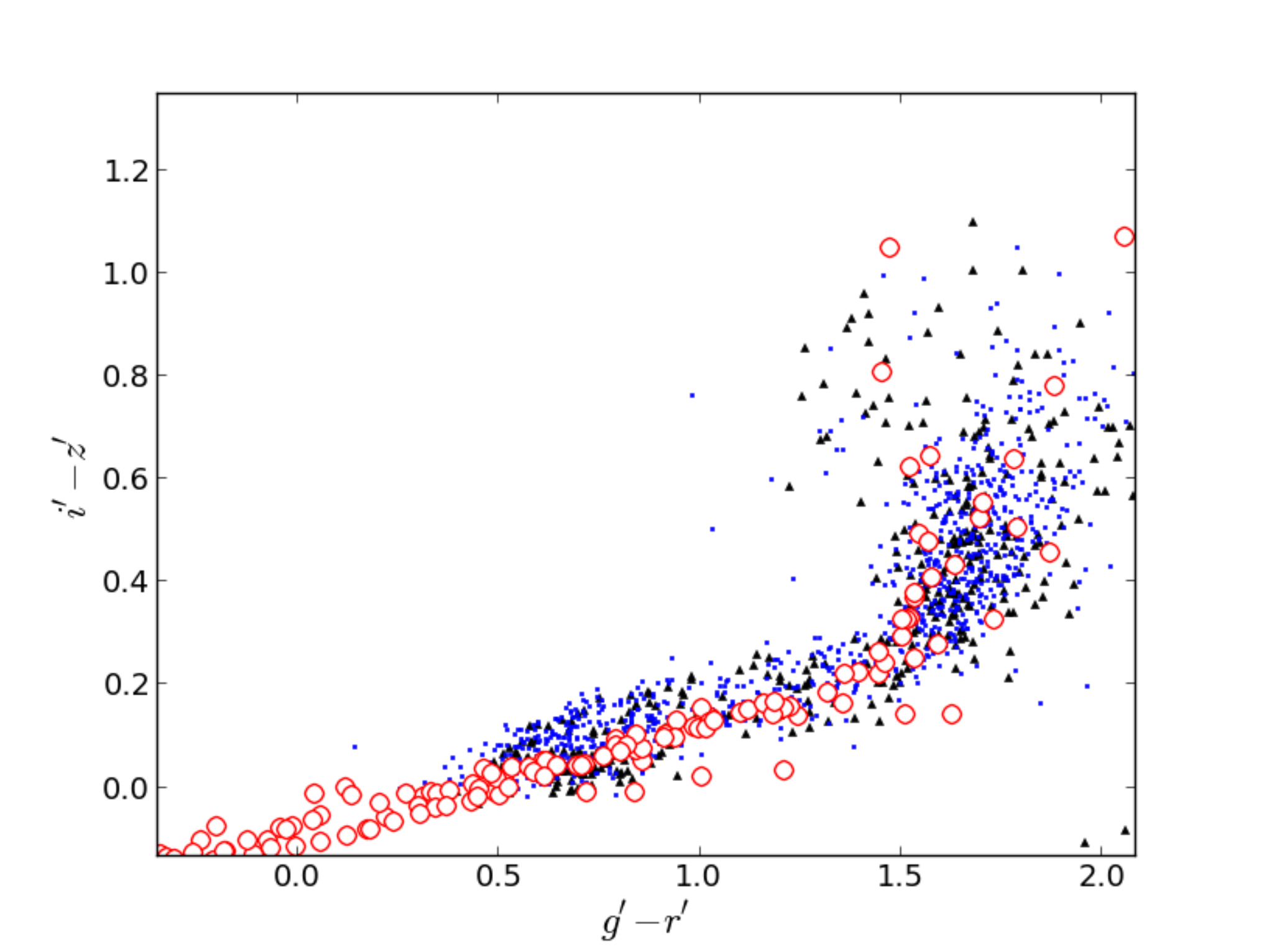}
}
	\newline
	\resizebox{\hsize}{!}{
       
\includegraphics[width=4.0cm]{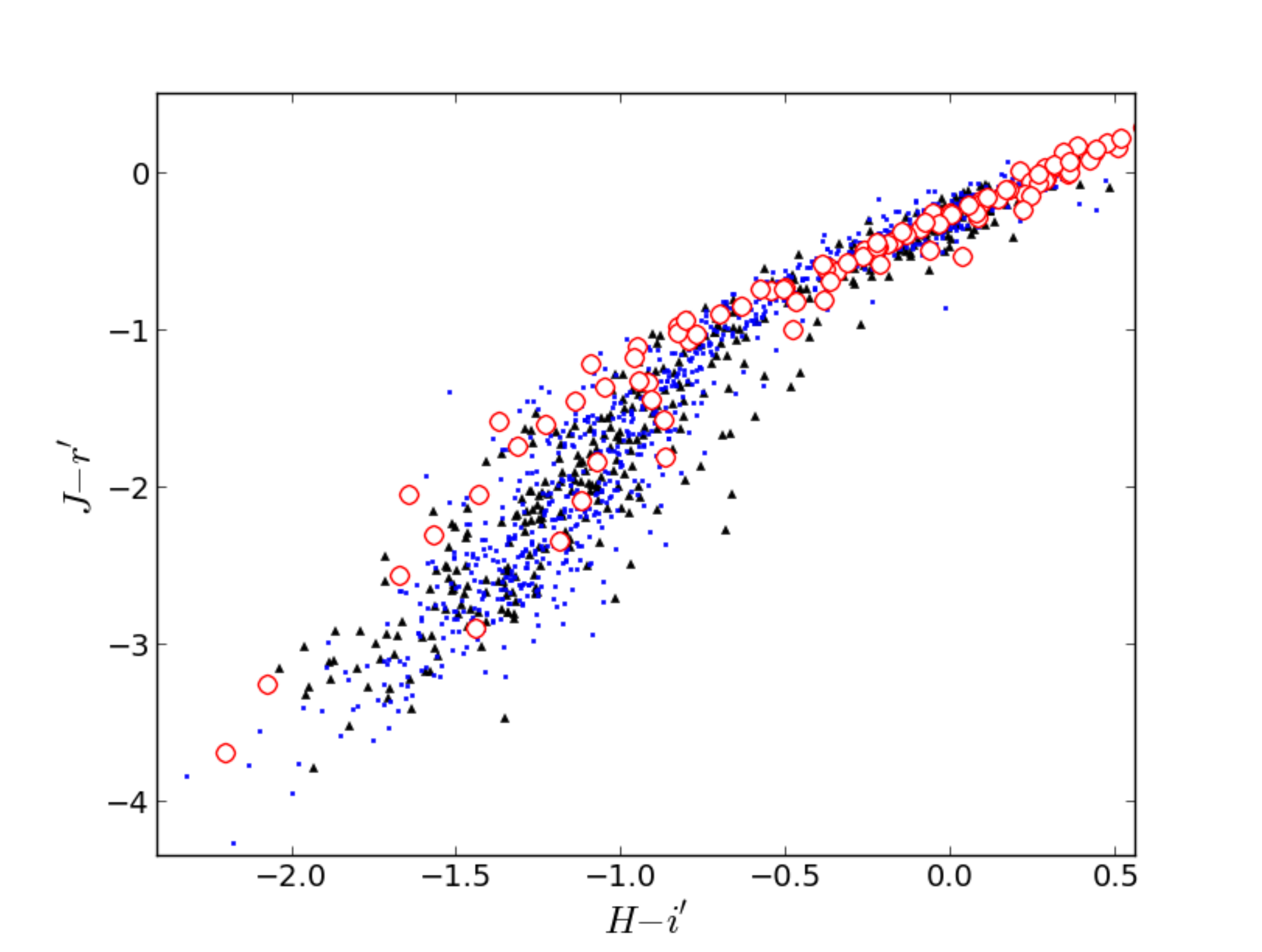}\includegraphics[width=4.0cm]{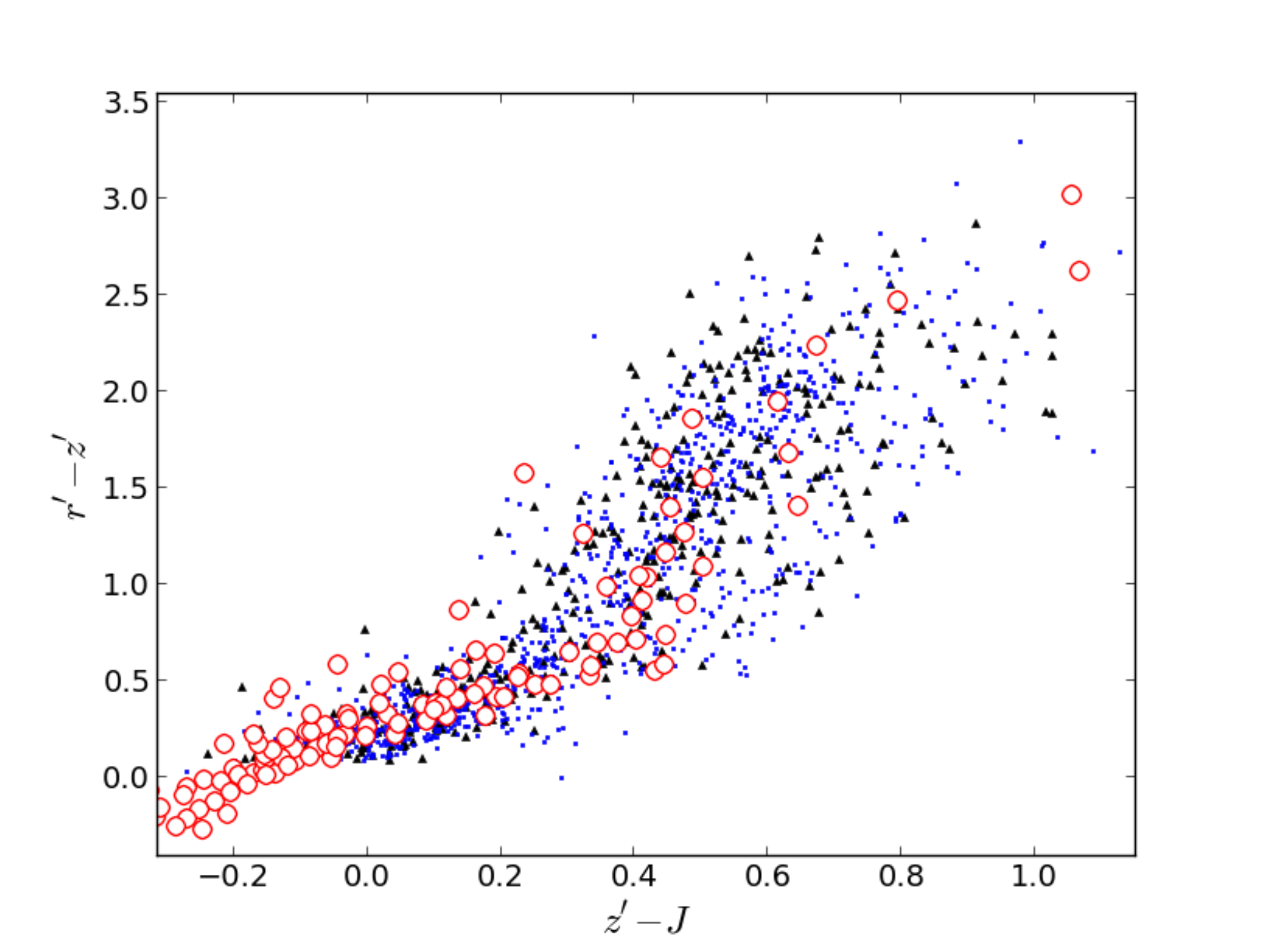}
}

        \caption{\emph{Color-color diagrams for stellar templates (open circles) and sources identified as stars (black triangles
	and blue squares) in
	the GROND sample. Offsets of $+0.2$, $+0.08$, $+0.08$, and $+0.1$ mag to the $g^\prime$, $r^\prime$, $i^\prime$, and
$z^\prime$ zeropoints  are applied, except in the case of direct SDSS calibration. The sources represented by black triangles
are SDSS calibrated, while the blue squares represent a zeropoint calibration.}}
        \label{fig:color_color_grond}
        \end{figure*}

To determine the accuracy of our galaxy-star separation, we compare a catalog of sources derived from nine GROND sub-fields
of
the COSMOS field with the COSMOS 30-band photometric catalog (COSMOS-30; \citealt{ILBERT2009}). COSMOS-30 has much
larger spectral coverage and higher sensitivity than the GROND observations, making it an ideal standard with which to
compare. We employ COSMOS-30 as both a measure of the quality of galaxy identification from GROND and to calculate the AGN-galaxy
two point correlation function. From the nine adjacent sub-fields observed by GROND (center pointing at R.A.=$150.0862$ deg,
Dec.=$2.3745$ deg), we identify $317$ total
candidate galaxies. Nine of these candidates are in a masked area in COSMOS-30, and a
further nine are categorized as stars in COSMOS-30. This corresponds to $2.9\%$ of galaxies that are
misclassified using our method, assuming COSMOS-30 has $100$\% accuracy in classification. Besides the masked objects, no
candidates were detected in GROND images without a corresponding source in COSMOS-30. The lack of spurious detections is likely a
result
from our stringent detection criteria discussed in \S \ref{subsection:creating_catalog}, namely that an object must have
independent detections in at least six filters.

We evaluate our photometric calibration by comparing the colors of sources that we classify as stars to the stellar 
templates of \citet{PICKLES1998} and \citet{BOHLIN1995}. In Fig. \ref{fig:color_color_grond}, we
present
color-color diagrams of our sources compared with those of stellar templates. On average, offsets of $+0.2$, $+0.08$, $+0.08$, and
$+0.1$ mag to the $g^\prime$, $r^\prime$, $i^\prime$, and $z^\prime$ zeropoints,
respectively, are required to match the colors of the stellar templates. In the case of direct SDSS calibration, no offset is
needed nor applied. After applying these offsets, the colors
of stars in our catalog are on average not offset from those of stellar templates, implying a reliable photometric calibration.
The somewhat higher scatter of our objects relative to the templates is due to a combination of uncertain dust
extinction towards foreground stars, binary systems, and photometric uncertainty.

\subsection{Measuring the two-point correlation}
After the source catalog is categorized, calculating a two-point GRB-galaxy correlation function is possible. Since our sample
consists of isolated fields with no overlap, we must calculate distance pair distributions on a field by field basis, later
combining them into a global correlation function. This technique
limits the angular sizes we can examine, and furthermore introduces systematic errors when the size scales approaches the size of 
individual images.
However, since we are specifically examining clustering at small scales, GROND's
$5.4^\prime \times 5.4^\prime$ images\footnote{$5.4^\prime \times 5.4^\prime$ in $g^\prime r^\prime i^\prime z^\prime$,
$10.3^\prime \times 10.3^\prime$ in $JHK$} have sufficient spatial coverage. We limit our our analysis to
angular distances $\le 120^{\prime\prime}$. This angular distance corresponds to $\sim ~600$ kpc at $z=0.4$, assuming a
$\Lambda$CDM concordance
cosmology with $h = 0.7$ and $\Omega_m=0.27$ \citep{NED_WRIGHT_COSMO_CALC}. 

Many of the afterglows in our sample have no measured
redshift. We assume that our sample of GRB afterglows has an underlying
redshift distribution that is consistent with the literature \citep{JAKOBSSON2012}. More specifically, the GRBs should be 
\emph{background} to the galaxies in the field. This latter assumption is likely justified, since the mean
redshift of GRBs is
$z\sim2$ and the mean redshift of galaxies with $K<19.3$ is $z\sim0.4$, as noted in \S \ref{section:sample}.

To measure the two-point correlation, we start by calculating the angular distance between the GRB optical afterglow positions (or
the XRT afterglow position circle,
in the case that no optical afterglow was
detected) and each galaxy. This provides a distribution of arclengths $n_{DD}$ for a given field. Next, we assign a new position
to each galaxy within that field based on values picked from a uniform random distribution. Any galaxies that are within two
stellar FWHM of a star or fall into an object mask are
re-assigned a position until they do not. This ensures that the observability of the isotropically distributed control
sample shares the same observational properties as the actual galaxies. We calculate the distances between each GRB-random
galaxy $n_{DR}$ and the random sample's mutual
separation $n_{RR}$. This process is repeated and averaged over a total of
$10^3$ times for each field. We then calculate the angular correlation function $w$ and its variance $\sigma$ using the Landy \&
Szalay
\citep{LANDY_SZALAY1993} estimator

\begin{eqnarray}
  w = \frac{DD-2DR+RR}{RR}, \\
  \sigma_{w}^2 = \frac{1+w}{n_{DD}} \label{equation:tpc},
\end{eqnarray}

where DD, DR, and RR are the normalized frequency distributions of $n_{DD}$, $n_{DR}$, and $n_{RR}$, respectively. We perform
the same procedure to measure $w_{galaxy-galaxy}$, $w_{AGN-galaxy}$, and $w_{random-random}$. Though it is trivial to show
analytically that $w$ is zero in the
case of isotropically distributed data, we explicitly perform the calculation using three randomly chosen coordinates
per field as a verification of our method.

\begin{figure}
	\resizebox{\hsize}{!}{
	\includegraphics[width=7cm]{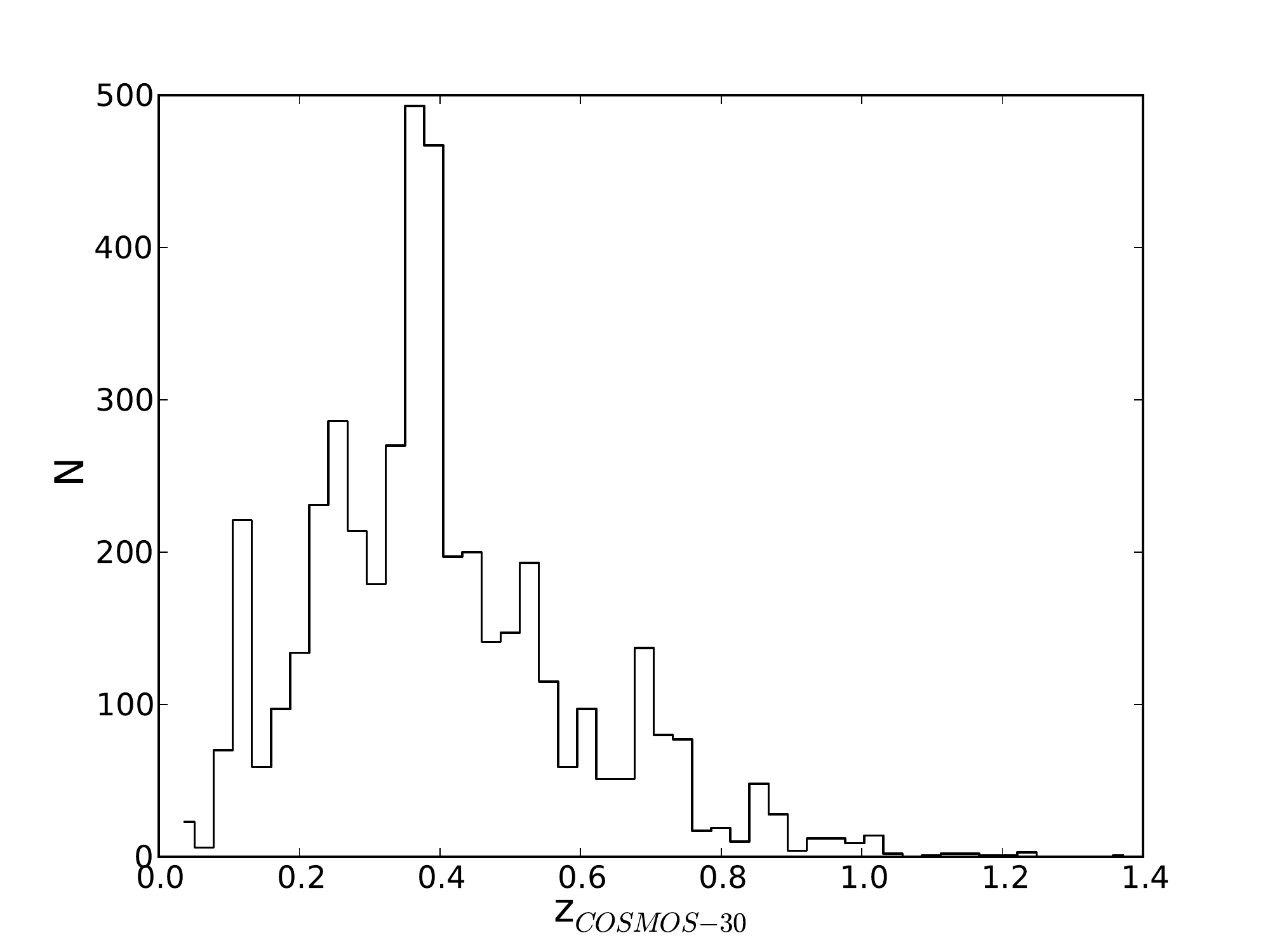}}
        \caption{\emph{Photometric redshift distribution of the $4481$ galaxies in COSMOS-30 that have $K_{\mathrm{AB}}<19.3$. The
mean and
median redshifts of the distribution are $0.387$ and $0.361$, respectively.}}
        \label{fig:zdistro}
        \end{figure}

We compute the correlation functions on both the GROND and COSMOS-30 sample. To approximate the same
selection criteria and completeness of our GROND sample, we remove from the COSMOS-30 catalog any sources dimmer than
K$_{\mathrm{AB}}=19.3$ as well as those sources whose photometry may be unreliable due to saturation or being within an object
mask (No flags in any optical filter). This leaves $4481$ galaxies and $2273$ stars.  The photometric redshift distribution of
these COSMOS-30
galaxies is presented in Fig. \ref{fig:zdistro}. The mean and median redshift of this sub-sample is $0.387$ and $0.361$,
respectively. For computing $w_{AGN-galaxy}$, we ensure that only background X-ray selected sources are included in the
calculation by imposing a $z>1.0$ redshift criterion \citep{SALVATO2011}. This criterion limits the catalog to $980$ sources.


\section{Results} \label{section:results}
\begin{figure} 
	\includegraphics[width=8.5cm]{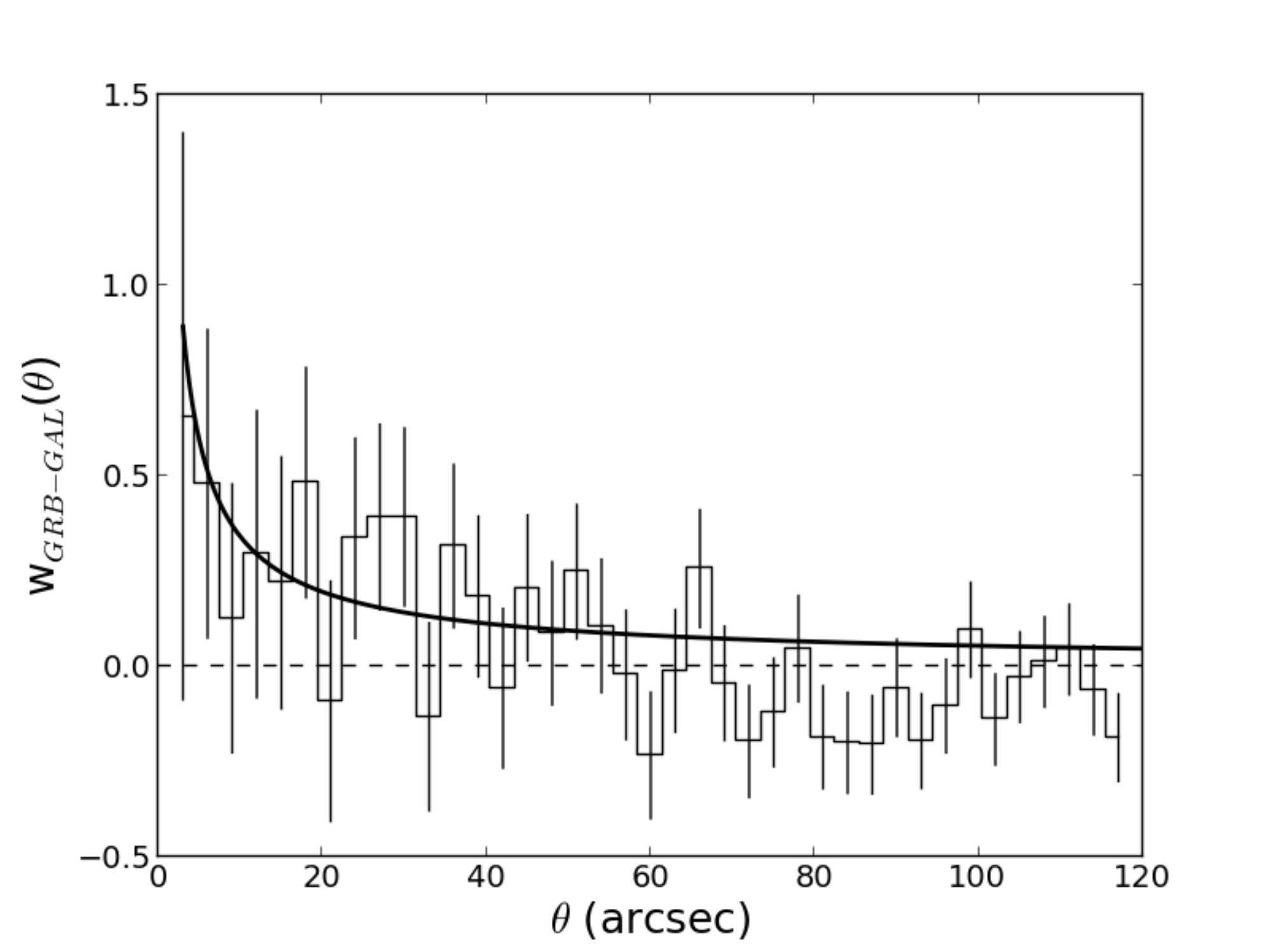}\newline
\includegraphics[width=8.5cm]{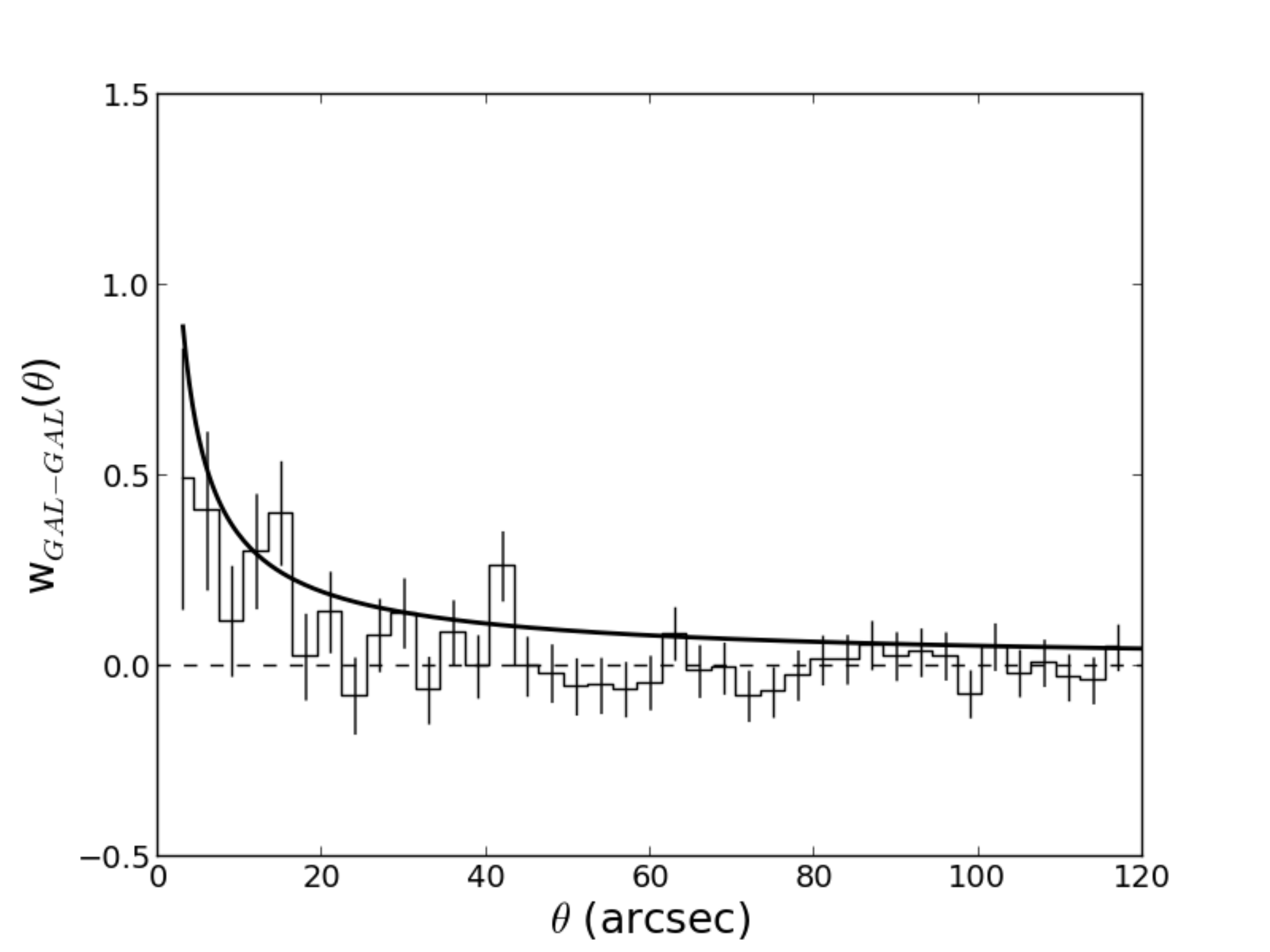}\newline
\includegraphics[width=8.5cm]{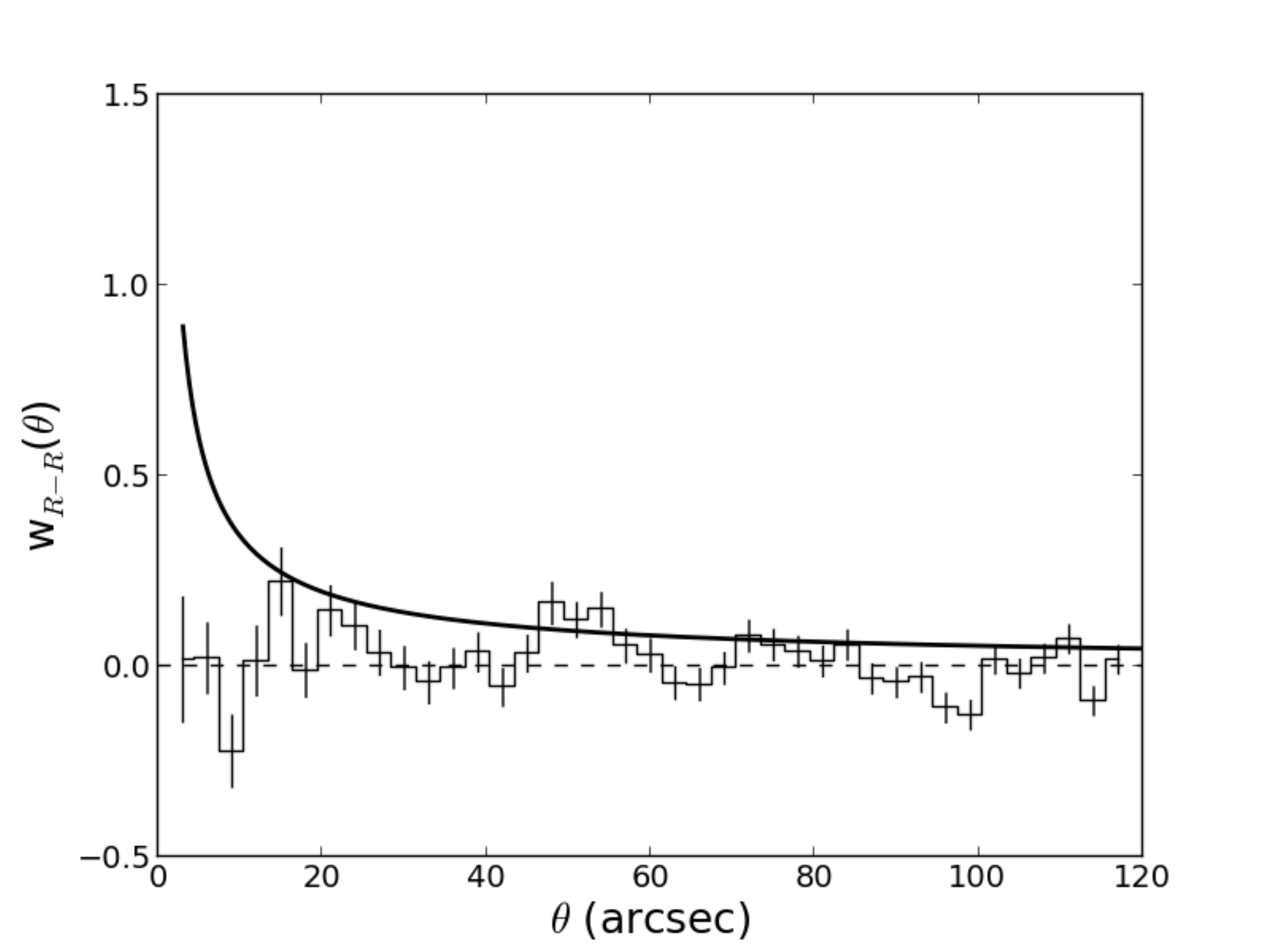}
        \caption{\emph{(Top to bottom) GRB-galaxy, galaxy-galaxy, and random-random two point correlation functions. The solid
black line represents the best fit power-law to the COSMOS-30 galaxy-galaxy two-point correlation function, as described in 
Fig. \ref{fig:tpcs_COSMOS} and in \S \ref{section:results}. The random-random two-point correlation function is, as
expected, consistent with zero.}}
        \label{fig:tpcs}
        \end{figure}

In Fig.\ref{fig:tpcs}, we present the two point GRB-galaxy, galaxy-galaxy, and random-random angular correlation
functions computed from the GRB sample. The two point correlation functions are self-consistent for all angular distance scales,
with
the exception of the
(as expected) null random-random function. A possible contribution to the first bin of the GRB-galaxy $w_{GRB-GAL}$ is likely due
to the detection of the host galaxy of the GRB, although the vast majority of known GRB hosts are fainter than our magnitude
limit. This corresponds to $3^{\prime\prime}$ in our sample, and it is entirely possibly to find GRB host galaxies at up to these
distances from the afterglow position and larger. Furthermore, these are also the angular distance scales within which
\citet{CHEN2009} find sources nearby sight-lines to GRBs with spectroscopically identified intervening MgII systems, albeit at
very faint magnitudes.

\begin{figure*}[h]
	\resizebox{\hsize}{!}{
	\includegraphics[width=5cm]{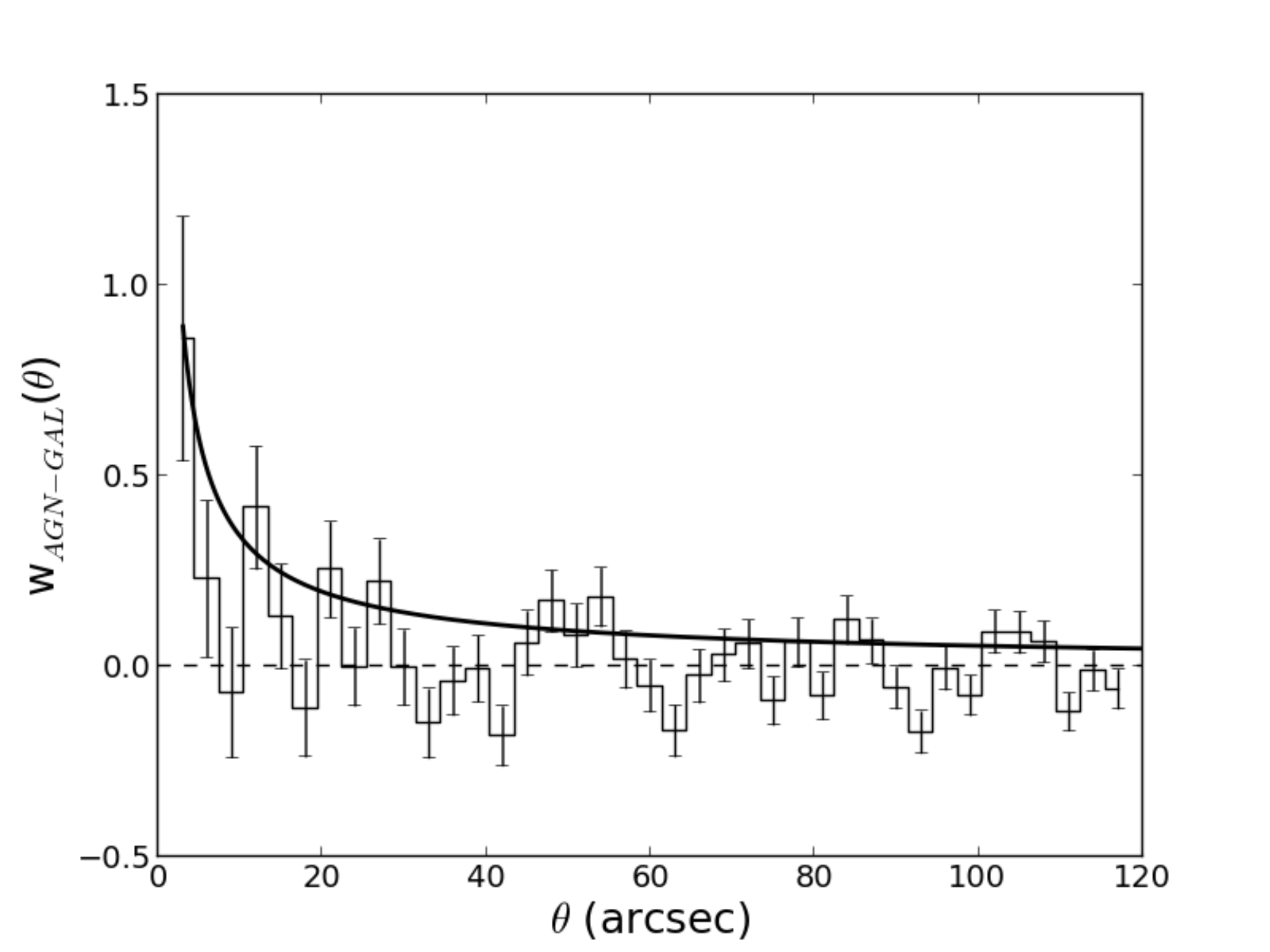}\includegraphics[width=5cm]{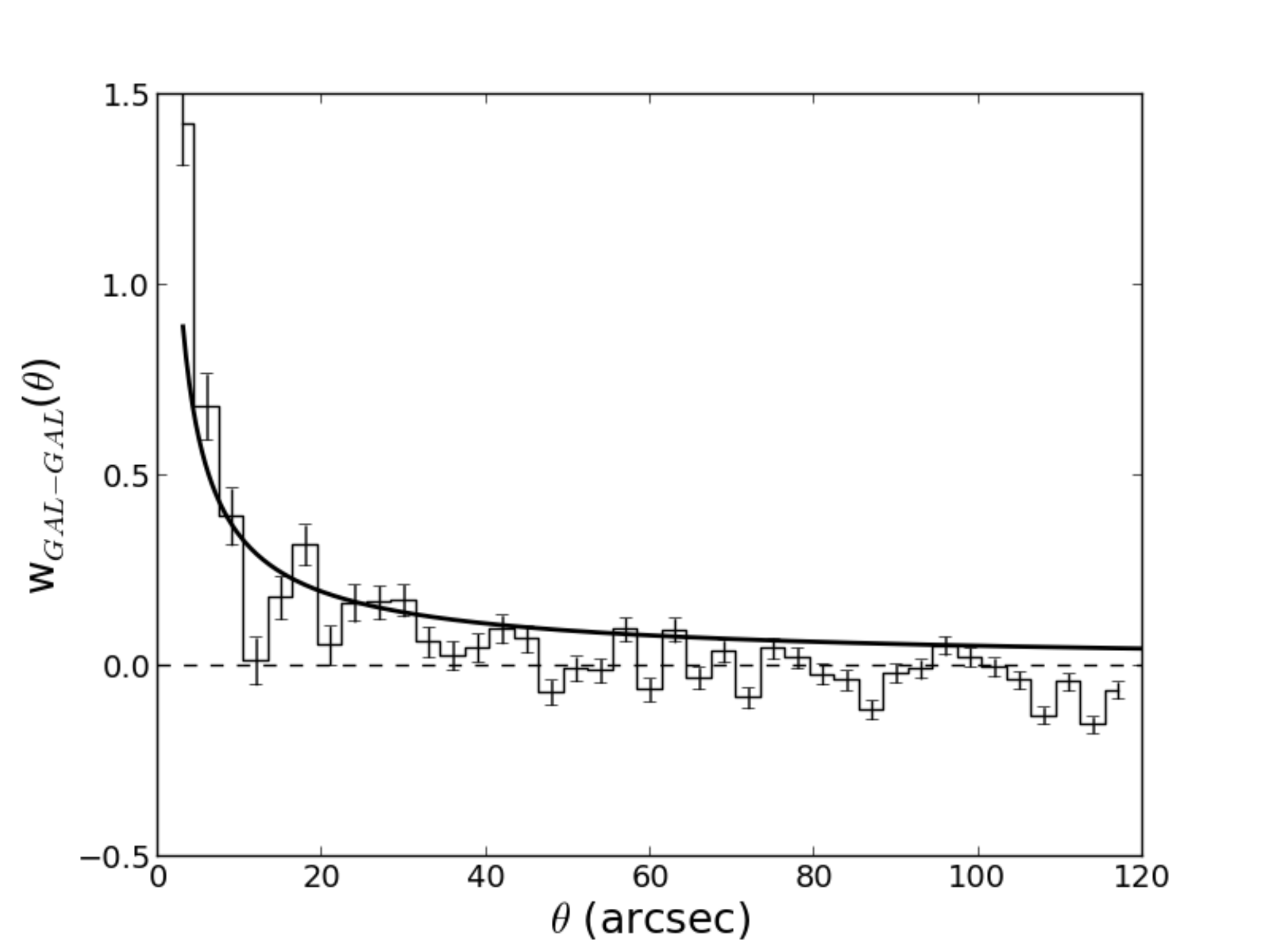}}
        \caption{\emph{(Left to right) AGN-galaxy and galaxy-galaxy two-point correlation functions computed from the COSMOS
photometric catalog. To approximate the GROND sample selection and completeness, only sources with $K_{\mathrm{AB}}<19.3$ are
included. The best fit power-law of the first $14$ bins ($0\le\theta\le42^{\prime\prime}$) of the galaxy-galaxy two-point
correlation function is represented by the solid black line. The same fit is used in both figures for comparison.}}
        \label{fig:tpcs_COSMOS}
        \end{figure*}

We also find clustering of galaxies within small angular distances of each other as
seen in the galaxy-galaxy correlation $w_{GAL-GAL}$. This is consistent with expectations from multiple galaxies occupying the
same massive halo
\citep{BERLIND_WEINBERG2002,LEE2006} and naturally from galaxy clusters. This signal is seen more clearly in the correlation
function as measured from the COSMOS-30
sample (Fig. \ref{fig:tpcs_COSMOS}), where we include only sources with K$_{\mathrm{AB}}\le19.3$ to approximate our sample's
completeness.
For the COSMOS-30 sample, the best fit power-law to the first
$14$ bins ($0\le\theta\le42^{\prime\prime}$) of the form $(\theta/\theta_0)^{-\delta}$ yields a correlation length $\theta_0 = 2.6
\pm 0.4$ arcseconds and slope $\delta = 0.8 \pm 0.2$ (statistical error only). This same power-law is presented in all two-point
correlation functions for comparison.

The AGN-galaxy two-point correlation function $w_{AGN-GAL}$ is also presented in Fig. \ref{fig:tpcs}.

\section{Discussion} \label{section:discussion}

A $w_{GRB-gal}$ that is consistent with $w_{GAL-GAL}$ and $w_{AGN-GAL}$ implies that there is no excess or unusual clustering
of galaxies
around GRB events, contrary to the expectation from observations of MgII absorption line systems. Though our data support this
null hypothesis, one must consider that the limiting
magnitude of our sample is
relatively shallow in comparison to the expected magnitudes of galaxies hosting MgII absorbers at the redshifts found in GRB
afterglow spectra. Indeed, the typical magnitudes of candidate absorption line
counterparts detected by \citet{SCHULZE2012} have $K_{\mathrm{AB}}>22$, and \citet{CHEN2009} find that the objects at angular
distances of
sources $1^{\prime\prime}-3^{\prime\prime}$ to GRB hosts with spectroscopically detected MgII typically have $H_{\mathrm{AB}}
\gtrsim
26$.

Although we expect that an overdensity of MgII absorbers corresponds to an overdensity of galaxies around
the line of sight, we are unfortunately limited to the very brightest and closest objects in our survey. A qualitative measure of
how much low number statistics affect $w$ is demonstrated by comparing our galaxy-galaxy correlation with that calculated from
COSMOS-30 (Figs. \ref{fig:tpcs}-\ref{fig:tpcs_COSMOS}). The correlation as measured by $3068$ galaxies in GROND fields only shows
a hint of clustering due to halo occupation of multiple galaxies and galaxy clusters, as discussed in \S~\ref{section:results}.
The signal is much more clear from the $4481$ galaxies with $K_{\mathrm{AB}}<19.3$ that we extract from COSMOS-30.

Clustering analysis of GRB afterglows and/or hosts have been performed by \citet{WANG_WEI2010,BORNANCINI2004}.
\citet{WANG_WEI2010} find no evidence of clustering between {\it SWIFT} XRT afterglows and {\it ROSAT} selected galaxy clusters.
The purpose of our survey is not to examine large scale anisotropies, but rather the small scale clustering implied by the MgII
excess. \citet{BORNANCINI2004} examined whether GRB host galaxies tend to reside in high density environments by calculating the
GRB-galaxy two point correlation for $19$ GRB hosts. They concluded that GRB host galaxies likely do not occur in over-dense
areas, although local cosmic variance could still significantly affect these results due to the small sample size. Furthermore,
choosing hosts over afterglows introduces an additional bias against intrinsically dim or reddened hosts. Constructing
a sample based on afterglow positions as we have done in this work removes this bias and increases the sample size significantly.
It is furthermore justified, since there is no a priori reason to link foreground absorption in the afterglow with the detection
or non-detection of a host galaxy.

The largest limitation to our survey stems from the fact that there is significant degeneracy between galactic/galactic and
galactic/stellar templates in GROND's optical regime for objects at $z<2$. For this reason, including NIR data is crucial to
distinguish point-like galaxies from stars. The sensitivity of GROND's $K$ channel is
typically three magnitudes shallower than its optical channels. Thus, we are forced to exclude a large number of sources from our
survey simply due to insufficient sensitivity. Including Spitzer/IRAC or HST/WFC3 data for a significant fraction of fields
observed by GROND would remove the largest constraint to our survey by providing much deeper sensitivity limits and/or wavelength
coverage, simultaneously increasing statistics and enabling the investigation of a much larger redshift space. Such surveys are
currently underway and are scheduled to be completed in the next years
\citep{LEVAN2009_HST_DARKBURST,LEVAN2009_HST_SNAPSHOT,PERLEY2012_SPITZER}. Combining these high
quality NIR data with GROND data would also provide for accurate photometric redshifts, enabling a measure of the two point
spatial correlation function.


\begin{acknowledgements}
We thank the anonymous referee for his or her helpful comments. We thank Sotoria Fotopoulou for insightful discussion regarding
photometric calibration techniques, and David Gruber for his projection plotting routine.

SK, DAK, and AR acknowledge support by DFG grant Kl 766/16-1. SS acknowledges support through project M.FE.A.Ext 00003 of the
MPG, and PS acknowledges support by DFG grant SA 2001/1-1.

TK acknowledges support by the European Commission under the Marie Curie Intra-European Fellowship Programme. The Dark Cosmology
Centre is funded by the Danish National Research Foundation.
\end{acknowledgements}


\newpage


\newpage
\begin{longtab}
\begin{longtable}{ccc}
\caption{\label{table:sample}Identification and positions of the GRBs in the final sample.}
\endfirsthead
GRB 	&  R.A. J2000 &  Dec. J2000  \\ \hline\hline
071112C & 02:36:50.93 & +28:22:16.68 \\
080212 & 15:24:35.42 & -22:44:29.70 \\
080330 & 11:17:04.51 & +30:37:23.48 \\
080408 & 07:38:39.59 & +33:18:14.90 \\
080413B & 21:44:34.67 & -19:58:52.40 \\
080413 & 19:09:11.74 & -27:40:40.30 \\
080514B & 21:31:22.69 & +00:42:28.90 \\
080520 & 18:40:46.30 & -54:59:31.00 \\
080523 & 01:23:11.70 & -64:01:51.50 \\
080605 & 17:28:30.05 & +04:00:55.97 \\
080710 & 00:33:05.67 & +19:30:05.69 \\
080916 & 22:25:06.20 & -57:01:22.90 \\
080928 & 06:20:16.82 & -55:11:59.30 \\
081008 & 18:39:49.89 & -57:25:52.80 \\
081109 & 22:03:09.86 & -54:42:41.04 \\
081118 & 05:30:22.18 & -43:18:05.30 \\
081121 & 05:57:06.08 & -60:36:09.94 \\
081228 & 02:37:50.89 & +30:51:09.10 \\
081221 & 01:03:10.19 & -24:32:52.20 \\
081230 & 02:29:19.53 & -25:08:51.72 \\
090102 & 08:32:38.10 & +33:11:45.30 \\
090123 & 00:27:08.74 & -23:30:04.00 \\
090205 & 14:43:38.65 & -27:51:10.70 \\
090323 & 12:42:50.29 & +17:03:11.60 \\
090401B & 06:20:21.10 & -08:58:19.35 \\
090423 & 09:55:33.29 & +18:08:58.00 \\
090424 & 12:38:05.09 & +16:50:14.75 \\
090426 & 12:36:18.04 & +32:59:09.24 \\
090509 & 16:05:39.01 & -28:23:59.64 \\
090516 & 09:13:02.59 & -11:51:14.90 \\
090519 & 09:29:07.00 & +00:10:49.10 \\
090530 & 11:57:40.51 & +26:35:38.40 \\
090812 & 23:32:48.54 & -10:36:17.60 \\
090814 & 15:58:26.35 & +25:37:52.42 \\
090902B & 17:39:45.41 & +27:19:27.10 \\
090926B & 03:05:13.94 & -39:00:22.20 \\
091018 & 02:08:44.61 & -57:32:53.70 \\
091029 & 04:00:42.60 & -55:57:19.84 \\
091109 & 20:37:01.80 & -44:09:29.60 \\
100219A & 10:16:48.50 & -12:34:00.50 \\
100414A & 12:48:26.93 & +08:41:34.40 \\
100518A & 20:19:09.33 & -24:33:16.56 \\
100621A & 21:01:13.11 & -51:06:22.46 \\
100902A & 03:14:30.96 & +30:58:45.23 \\
101023A & 21:11:51.23 & -65:23:15.61 \\
101219B & 00:48:55.34 & -34:33:59.26 \\
110128A & 12:55:35.10 & +28:03:54.10 \\
110206A & 06:09:20.04 & -58:48:24.91 \\
110312A & 10:29:55.47 & -05:15:45.20 \\
110407A & 12:24:07.49 & +15:42:42.16 \\
110709B & 10:58:37.11 & -23:27:16.76 \\
110721A & 22:14:38.19 & -38:35:35.70 \\
110818A & 21:09:20.89 & -63:58:51.80 \\
111008A & 04:01:48.22 & -32:42:34.09 \\
111107A & 08:37:54.65 & -66:31:12.40 \\
111129A & 20:29:44.14 & -52:42:46.48 \\
111209A & 00:57:22.70 & -46:48:05.00 \\
111211A & 10:12:21.70 & +11:12:30.00 \\
111212A & 20:41:43.52 & -68:36:45.00 \\
111228A & 10:00:16.01 & +18:17:51.80 \\
111229A & 05:05:08.84 & -84:42:38.70 \\
120119A & 08:00:06.94 & -09:04:53.83 \\
120211A & 05:51:00.89 & -24:46:30.79 \\
120302A & 08:09:35.54 & +29:37:41.05 \\
120311A & 18:12:22.16 & +14:17:46.30 \\
120320A & 14:10:04.30 & +08:41:47.26 \\
120404A & 15:40:02.29 & +12:53:06.29 \\
120422A & 09:07:38.42 & +14:01:07.36 \\
121024A & 04:41:53.30 & -12:17:26.48 \\
121027A & 04:14:23.45 & -58:49:47.17 \\
121217A & 10:14:50.51 & -62:21:03.28 \\
121229A & 12:40:24.29 & -50:35:39.48 \\
\hline
\end{longtable}
\end{longtab}


\bibliographystyle{aa}
\bibliography{main_bibliography} 

\end{document}